# Simulation System for the Wendelstein 7-X Safety Control System

J. Schacht, A. Wölk, S. Pingel, U. Herbst, D. Naujoks and the W7-X Team

*Abstract*—The Wendelstein 7-X (W7-X) Safety Instrumented System (SIS) ensures personal safety and investment protection. The development and implementation of the SIS are based on the international safety standard for the process industry sector, IEC 61511. The SIS exhibits a distributed and hierarchical organized architecture consisting of a central Safety System (cSS) on the top and many local Safety Systems (lSS) at the bottom. Each technical component or diagnostic system potentially hazardous for the staff or for the device is equipped with an lSS. The cSS is part of the central control system of W7-X. Whereas the lSSs are responsible for the safety of each individual component, the cSS ensures safety of the whole W7-X device.
For every operation phase of the W7-X experiment hard- and software updates for the SIS are mandatory. New components with additional lSS functionality and additional safety signals have to be integrated. Already established safety functions must be adapted and new safety functions have to be integrated into the cSS. Finally, the safety programs of the central and local safety systems have to be verified for every development stage and validated against the safety requirement specification.
This contribution focuses on the application of a model based simulation system for the whole SIS of W7-X. A brief introduction into the development process of the SIS and its technical realization will be give followed by a description of the design and implementation of the SIS simulation system using the framework SIMIT (Siemens). Finally, first application experiences of this simulation system for the preparation of the SIS for the upcoming operation phase OP 1.2b of W7-X will be discussed.

## I. INTRODUCTION

THE fusion experiment Wendelstein 7-X (W7-X) is a superconducting stellarator with a capability for steady state plasma operation. The first operational phase of W7-X started in December 2015 [1]. Meanwhile two operation phases had been conducted with very good results with respect to availability and performance of the W7-X systems during operation. Furthermore, very good experimental results could be obtained [2].

An important aspect of W7-X device operation is to prevent any personnel or machine hazards. A Safety Instrumented System (SIS) was designed and implemented based on the EN/IEC 61511 standard for the "functional safety – safety instrumented systems for the process industry sector". This standard is the most appropriate one, since W7-X does not fall under the category of a nuclear plant.

The requirements on safety instrumented functions (SIF) were derived from the safety concept for the main device and infrastructural environment as well as from the risk assessments of all attached components. The functions have been distinguished in those belonging to personnel safety and those dedicated for the protection of the investment, i.e. device safety.

Every new operation phase generates new or modified requirements on safety functions. For example, new technical or diagnostic systems have to be integrated into the W7-X control systems or the parameter space of operation will be broadened.

During the completion phases of W7-X the SIS has to be modified. This modification comprises both the implementation of new safety signal interfaces as well as the programming of new safety instrumented functions (SIF). All SIFs of an operation phase are described in a safety requirement specification (SRS). Every commissioning of W7-X required a well validated safety control system according to the required safety standards. Therefore, different tests and validation steps have to beconducted, before the SIS is approved for the next operation phase. At the end of the software implementation tasks, module tests of the individual function blocks have to be performed and documented. Also the sensors, actuators and the correct cabling have to be checked and documented. The final tests are the validation of the SIFs against the SRS.

Since the time period for a W7-X completion phase is often very short, it is mandatory to optimize the processes for software development and testing. The Control and Data acquisition group (CoDa group) decided to use a model based simulation framework based on the SIMIT framework (Siemens) for supporting the software developments and for integration tests of the SIS for the commissioning phase CP2.1b.

After a short description of the SIS the framework SIMIT and its application for the SIS of W7-X will be described.

## II. ARCHITECTURE OF W7-X SIS

Figure 1 shows the overall architecture of the control system, which is a three-tier hierarchical system. The safety system, which is described here solely, acts on different safety

J. Schacht, A. Wölk, S. Pingel, U. Herbst, D. Naujoks are with the Institute for Plasma Physics, Wendelsteinstraße 1, Greifswald, Germany, D-17491, telephone: +049(0)3834-882761, (e-mail: joerg.schacht@ipp.mpg.de).

layers and gives operation permits to the operations management and subsequent to the plasma control system. Regarding the architecture, it had been decided to choose the same distributed control system as for the operations management, a star-like topology with the central safety system (cSS) and the local safety systems (lSS) belonging to the components such as the heating systems, auxiliary systems (e.g. gas inlet) and plasma diagnostics.

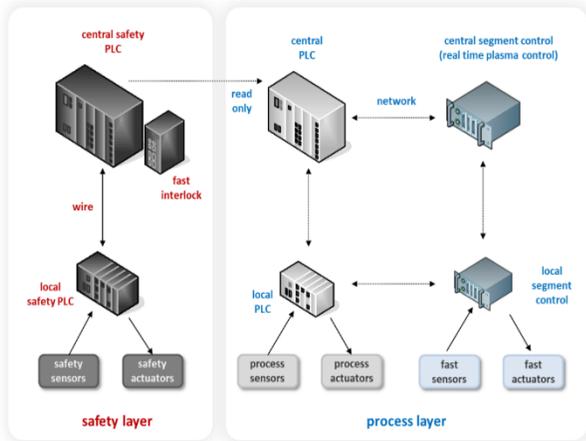

Figure 1. Architecture of the Wendelstein 7-X control system.

The structure of the SIS software is shown in figure 2.

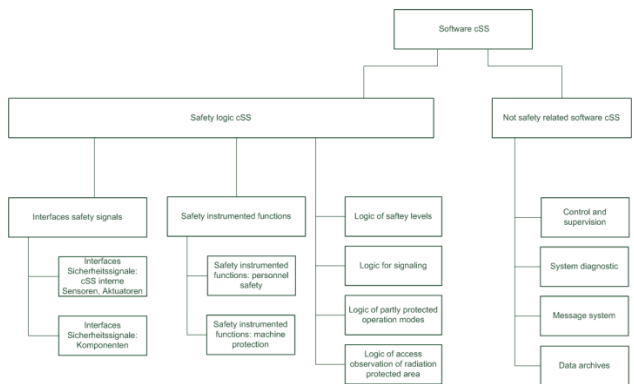

Figure 2. Structure of SIS software.

### III. SIS SIMULATION SYSTEM

The simulation platform SIMIT (company Siemens, Germany) is applicable for a simulation of devices and complex plants [3]. The main advantages of this simulation platform are:
- Developing and testing of the automation software before commissioning,
- Emulation of signals, device and process levels,
- Graphical user interface for an easy handling, and
- Utilization as operator training system.

The application of the model based simulation tool SIMIT allows to test the SIS software without a running W7-X device. The test environment consists of the same PLC hardware for the cSS as the original system.

The behavior of the local safety systems of the test system is virtualized. The PLC software version of the cSS test system depends on the test task which will be performed. For failure detecting of the actual productive SIS software version this version must be loaded into the PLCs of the test system. The second application comprises module and integration tests of new software versions for upcoming operation phases. These tests can be performed in parallel to an operating W7-X SIS without any interference.

The architecture of the SIS test system is shown in figure 3.

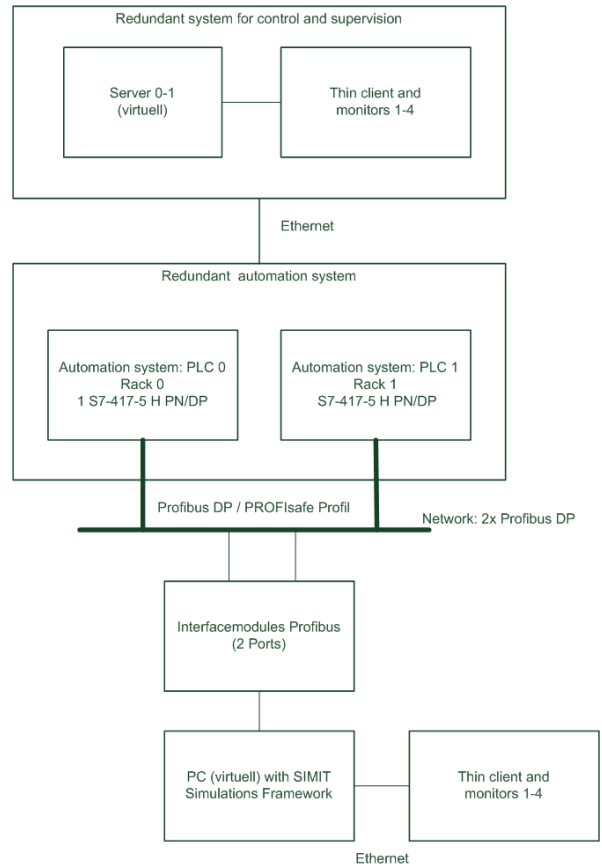

Figure 3. Architecture of W7-X SIS test systems.

The SIS software under test runs on a redundant automation system based on 2 PLC S7-417 with fail-safe and high reliability capabilities. The hardware and the network configuration of the PLC are identical to the W7-X SIS except the network addresses.

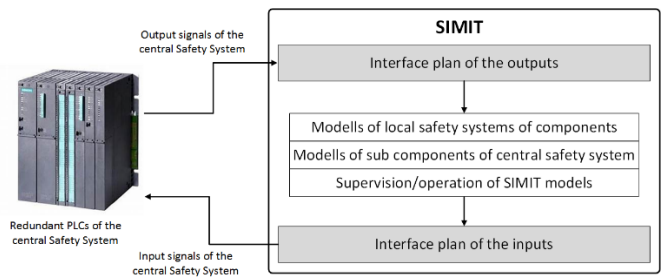

Figure 4. Signal interfaces and model types of the SIS simulation program.

Fig. 4 shows the coupling of the PLC safety program under test with the simulation program running on the computer, which is equipped with the SIMIT simulation tool SIMIT V9.0 and the simulation software. The signal transfer between the PLCs and the simulation program is established by using a Profibus safe network data connection. A special simulation unit for Profibus (Siemens, Germany) emulates all signal connections between the PLCs and decentral periphery devices of the SIS. The so-called interface planes for input and output signals realizes the transfer of signals between the PLC program and the SIMIT models.

The simulation software comprises the models for the local safety components (e.g. for models the power supplies of the superconducting magnets), models of hardware subcomponents of the central safety system (e.g. emergency stop switches, signalization, access control system), and the models for supervision and operation of the SIMIT models. The simulation models can be controlled and observed via graphical operator panels. Furthermore, complex control sequences of a simulation run can be defined by programming of macros.

## IV. REALIZATION AND RESULTS

In the design phase of the SIS simulation platform, all planned simulation models were described in a specification by using the SysML modeling language [4]. Especially the diagram types 'finite state diagrams' and 'activity diagrams' are suitable for this purpose. Figure 5 depicts a SysML finite state model of the coil current of the superconducting magnet system.

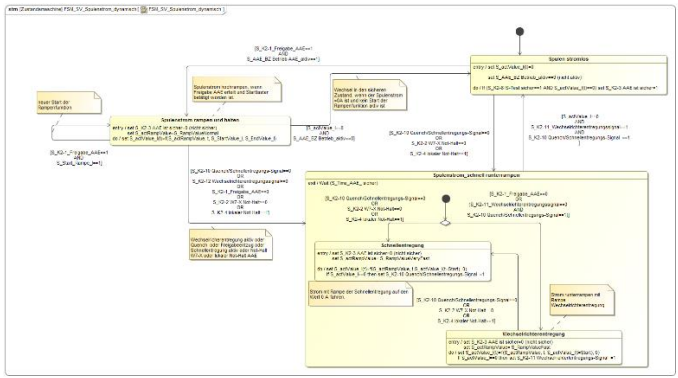

Figure 5. SysML finite state machine model for modeling of coil current.

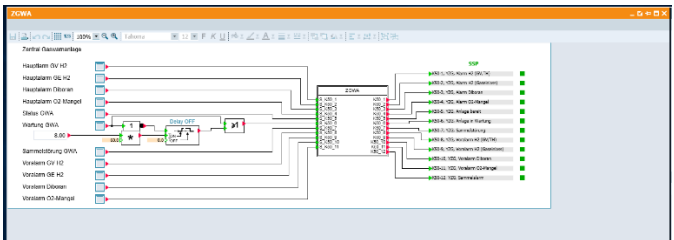

Figure 6. SIMIT program chart for modelling of central gas warning system.

Program charts are used to realize specified models in SIMIT . SIMIT offers a large library with standard control elements. In addition, a more complex and specific behavior can be encapsulated by defining user specific control elements. Fig. 6 shoes a specific control element for the central gas warning system as an example for a user specific control element.

For the control of simulation variables, the user can choose different standard tools of SIMIT, e.g. the tables of variables for showing and modifying variables. More comfortable are the user defined control panels, which allows to combines graphical elements, text boxes, output of values, and placing of control element like switches and knops for various user inputs.

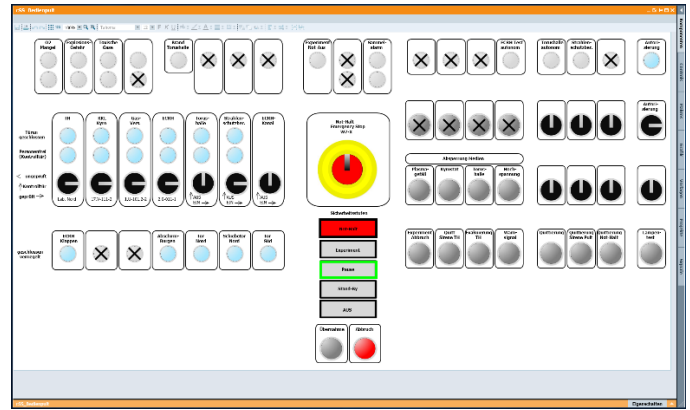

Figure 7. User specific SIMIT control panel for emulation of the cSS control console.

A very realistic implementation of the cSS control console in SIMIT as a control panel is shown in Fig.7. The user can operate this panel in the same way as the real cSS control console.

More than 2000 input and output variables of the safety program are processed in the actual simulation software. Simulation models have been prepared and implemented for all technical components and diagnostics which have a safety interface to the cSS. In addition, important hardware subcomponents of the cSS, like the gas warning system, the access system of the radiation protection area, the emergency stop system, and the signalization are incorporated in the simulation software.

The hardware setup for the simulation system was entirely separated from the hardware of the W7-X safety system. The PLCs, the network devices and the Profibus coupling devices have been installed permanently in an electrical cabinet. The software version of the safety program of the PLCs corresponds to the planned simulations.

For the preparation of the cSS for a new operation phase, the PLC safety software development and the tests can be performed by using the combined engineering and simulation platform. Furthermore, the simulation software must be modified due to the upgrade of the established models or due to the implementation of new models. During a running operation phase, the PLC software for the simulation platform is the same as the software of the operating safety system. This allows a comfortable simulation and debugging of failure situations, which had been detected during operation of the operating safety system.

The working place of the simulation platform consist of a PC, running the simulation tool SIMIT V9.0 and the simulation software, and four 19" monitors as graphical user interface. The

WinCC visualization program of the PLCs runs on virtualized hardware and is visualized with another four 19" monitors.

Finally, a separate operator station with two 19" monitors is used as an engineering station allowing modifications of the PCS-7 safety program.

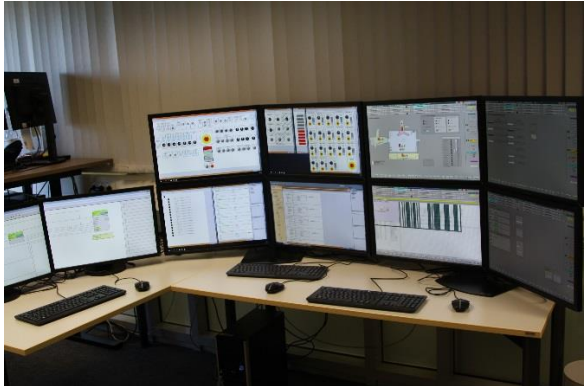

Figure 8. Workplace for SIMIT simulation platform of cSS.

Fig. 8 shows a total view on the workplace of the simulation platform.

The design, installation and commissioning of the hardware of the simulation platform were performed without any major problems. The CoDa (CoDa: Control and Data acquisition) group was supported by the external company BST GmbH, Germany during the development of the simulation software. Due to the large experiences of BST GmbH with simulation projects for industry plants, the specified models could be realized in a very short time of less than 4 months.

By using the simulation platform during the software development and software integration tests for operation phase 1.2b, an early detection and correction of configuration and software errors were possible. It is worth to mention that during the validation of the cSS program for OP1.2b no serious error was detected. This made the commissioning of the cSS for OP1.2b possible in time. The simulation platform provides the possibility of safety software development parallel to W7-X operation.

The simulation platform will be used in the near future for the education of cSS operators and other staff members. Another application will be the debugging of real failure scenarios, which could occur during W7-X operation.

## V. STATUS AND OUTLOOK

The fusion experiment W7-X has successfully accomplished the commissioning and two operation phases (OP1., OP1.2a). For more efficient software development, module and integration tests, a SIMIT based test system for the SIS was established and commissioned. Actually, about 1300 fail-safe input signals and 700 fail-safe output signals of the cSS can be controlled. SIMIT models for the behavior modeling of all important functions of the cSS hardware modules and lSS SIFs have been designed and developed. The software tests using the SIS test system were performed successfully end of February 2018, after completion of the SIS software development for the upcoming operation phase OP1.2b.

## VI. ACKNOWLEDGMENTS

The authors are grateful to S. Katzki (Buck Systemtechnik GmbH, Germany) for implementing the simulation models with the SIMIT simulation tool (Fa, Siemens), I. Müller for setting up reliable network communication, and D. Aßmus and M. Wallis for implementing the required electro-technical setup for the simulation system.

This work has been carried out within the framework of the EUROfusion Consortium and has received funding from the Euratom research and training programs 2014-2018 under grant agreement No 633053. The views and opinions expressed herein do not necessarily reflect those of the European Commission.